\documentclass[twocolumn,showpacs,preprintnumbers,amsmath,amssymb]{revtex4}
\usepackage{graphicx}
\usepackage{dcolumn}
\usepackage{bm}

\begin{document}

\title{Universal Scaling of Pinning Effect on Hall Anomaly
near the\\ Vortex Glass Transition and Doping Dependence Problems of
Superconductors}

\author{L. He, X. Hu, L. Yin, H. Y. Xu, X. L. Xu, J. D. Guo, C. Y. Li, and D. L. Yin}

\email{yindl@pku.edu.cn}

\affiliation{School of Physics, Peking University, Beijing 100871,
China}
\date{\today}

\begin{abstract}
We find universal scaling relations of the pinning effect on the
Hall resistivity $\rho_{xy}$ and Hall angle $\theta_{H}$.
Considering the extended power law form of $\rho_{xx}$ and the
microscopic analysis of $\sigma_{xy}$, we obtain unified $\rho_{xy}$
equations for superconductors with and without double sign reversal.
These equations reasonably explain the striking universality in
doping dependence found by Nagoaka et al., which contradicts the
prediction of the time dependent Ginzburg-Landau equation based on
s-wave coupling theory [PRL {\bf{80}},3594 (1998)]. A full
comparison of experiment with prediction from theoretical models is
proposed.
\\
\end{abstract}

\pacs{74.25.Sv, 74.25.Fy, 74.25.Ha, 74.20.De} \maketitle

{\it Introduction}---One of the most puzzling and controversial
phenomena in the field of vortex physics of high temperature
superconductors(HTS) is the sign reversal, including single, double
and triple reversal that has been observed in the Hall effects in
the superconducting state of most HTS and some conventional
superconductors\cite{Hagen,2}. It contradicts the well-known
conventional theories of flux motion\cite{Bardeen,Nozieres} which
predict that the Hall effect stems from quasi-normal core and hence
should have the same sign as in the normal state. Another puzzling
issue is the power law scaling relation between the Hall resistivity
$\rho_{xy}$ and longitudinal resistivity $\rho_{xx}$
\cite{Luo,Kang}. Experiments show that the Hall effect at least
depends on two factors: the doping level \cite{Nagaoka} and the
vortex pinning \cite{Luo,Woltgens, Kopnin}. Vinokur, Geshkenbein,
Feigel'man and Blatter (VGFB) calculated the effects of flux pinning
on Hall resistivity $\rho_{xy}$ and show
$\rho_{xy}\propto\rho_{xx}^{\beta}$ with $\beta=2$ for small Hall
angles $\theta_{H}$ \cite{Vinokur}. However, the scaling exponent
was not always observed universal and varied from $\beta\approx 2$
to $\beta\approx 1$ depending on the applied magnetic field and
concentration of defects \cite{Kang}. Using a simple model for
pinning potential, Kopnin and Vinokur show \cite{Kopnin} that the
magnitude of the vortex contribution to the Hall voltage decreases
with increase in the pinning strength. $\rho_{xy}$ scales as
$\rho_{xx}^2$ only for weak pinning, strong enough pinning can break
the scaling relation and can even result in a sign reversal of
$\rho_{xy}$. The effect of pinning on the Hall anomaly is not well
understood and there are several contradicting views, so more
studies are needed to investigate the Hall resistivity in the flux
creep regime. Based on the hypothesis of a vortex glass transition
\cite{Fisher}, and the time dependent Ginzberg-Landau (TDGL) theory,
Dorsey and Fisher \cite{Dorsey} reasonably explained the puzzling
scaling behavior $\rho_{xy}\propto\rho_{xx}(T)^{\beta}$ with
$\beta=1.7\pm0.2$ as observed by Luo {\it et al.} \cite{Luo}. The
Hall effect itself was attributed to particle-hole asymmetry and the
exponent $\beta$ was related to a specially chosen particle-hole
asymmetry exponent $\lambda\simeq3$. They predicted \textquoteleft
\textquoteleft The nonlinear Hall electric field $E_y$ should
exhibit universal scaling, and right at the transition, should
vanish with a universal power of the current density $J_x$". This
prediction has been verified in the experiment by W\"{o}ltgens {\it
et al.}\cite{Woltgens} in the case of YBa$_2$Cu$_3$O$_{7-\delta}$
(YBCO) with a Hall-related exponent $\lambda=3.4\pm0.3$.
Furthermore, they found wide range scaling behavior
$\rho_{xy}=A\rho_{xx}(J_x,T)^{2.0\pm0.2}$ in consistency with the
model proposed by VGFB \cite{Vinokur}. Based on the normal core
model of Bardeen-Stephen(B--S) \cite{Bardeen} and Nozi\`{e}res-Vinen
(N--V) \cite{Nozieres}, Wang, Dong and Ting (WDT) developed a
unified theory to explain the sign reversal and scaling behavior of
Hall effect \cite{WDT}. Xu {\it et al.} \cite{Xu} resolved the
nontrivial problem of determining the positive scaling function
between the average pinning force $\langle{\bf F}_p\rangle_t$ and
vortex velocity $v_L$ and found a Hall resistivity equation which
agrees with the experimental data of YBCO with single sign change in
$\rho_{xy}$ but fails in describing the behavior of HTS with double
sign reversal, e.g.,HgBa$_2$CaCu$_2$O$_{6+\delta}$ (HBCCO). It is
the purpose of the presented letter to report the universal scaling
behavior of the pinning effect on the Hall resistivity and Hall
angle in superconductors both with and without double sign reversal,
and to propose a full comparison of Hall anomaly experiment with
predictions of theoretical models.\\
\\

\begin{figure*}
\includegraphics[width=0.8\linewidth]{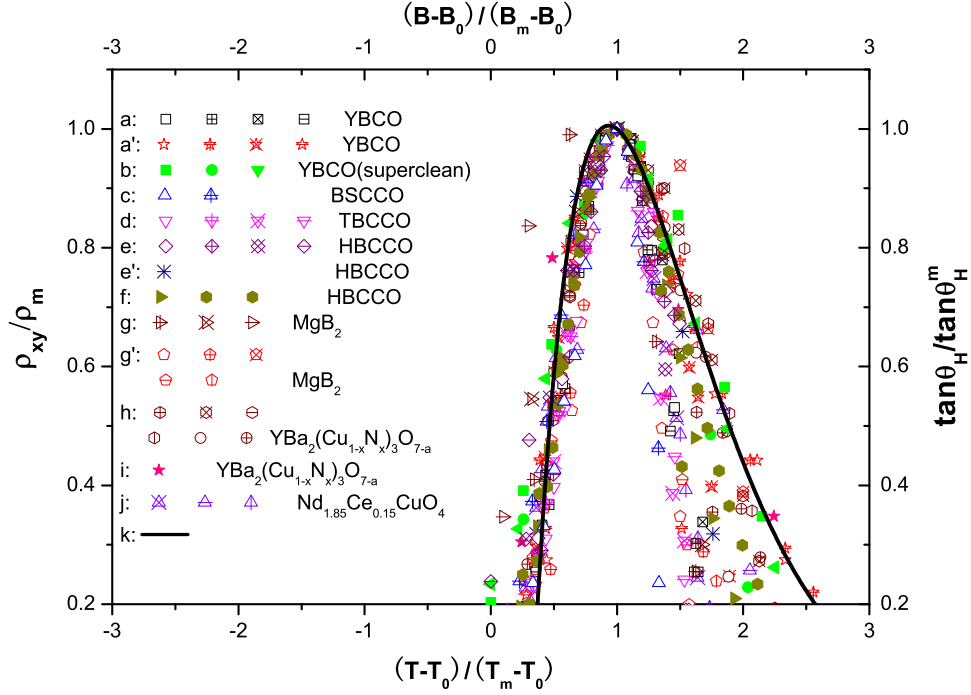}
\vspace{-7mm} \caption{Universal scaling of Hall resistivity
$\rho_{xy}$ and Hall angle $\theta_{H}$: a,c,d,e,g,j represent
scaling of the anomalous $\rho_{xy}\sim T$ curves of different kinds
of superconductors in the form of
$\log_{10}(\rho_{xy}/\rho_m)~\text{vs}~\log_{10}((T-T_0)/(T_m-T_0))$
; b,f represent scaling of the anomalous $\tan\theta_{H}\sim T$
curves in the form of
$\log_{10}(\tan\theta_H/\tan\theta_H^m)~\text{vs}~\log_{10}((T-T_0)/(T_m-T_0))$;
a',e',g',h represent scaling of the anomalous $\rho_{xy}\sim B$
curves of different kinds of superconductors in the form of
$\log_{10}(\rho_{xy}/\rho_m)~\text{vs}~\log_{10}((B-B_0)/(B_m-B_0))$;
i represent scaling of the anomalous $\tan\theta_{H}\sim B$ curves
in the form of
$\log_{10}(\tan\theta_H/\tan\theta_H^m)~\text{vs}~\log_{10}((B-B_0)/(B_m-B_0))$.\\
(a) represents YBa$_2$Cu$_3$O$_7$ \cite{Hagen}; (a') represents
YBa$_2$Cu$_3$O$_7$ \cite{Chien}; (b) represents superclean
YBa$_2$Cu$_3$O$_{6+y}$ \cite{Harris}; (c) represents
Bi$_2$Si$_2$CaCu$_2$O$_{8+x}$ \cite{Ri}; (d) represents
Tl$_2$Ba$_2$CaCu$_2$O$_8$ \cite{HagenLobb}; (e) represents
HgBa$_2$CaCu$_2$O$_{6+\delta}$ before and after irradiation
\cite{Kang}; (e') represents HgBa$_2$CaCu$_2$O$_{8+\delta}$
\cite{KK}; (f) represents HgBa$_2$CaCu$_2$O$_{6}$ \cite{KangChen};
(g,g') represents MgB$_2$ \cite{Jin}; (h) represents Ni-doped single
crystal of YBa$_2$(Cu$_{1-x}$N$_{ix}$)$_3$O$_{7-\delta}$ with
$x=0.005, 0.015, 0.035$ \cite{Kim}; (i) represents Ni-doped single
crystal of YBa$_2$(Cu$_{1-x}$N$_{ix}$)$_3$O$_{7-\delta}$ with $x=
0.035$ \cite{Kim}; (j) represent the n-type superconductor
Nd$_{1.85}$Ce$_{0.15}$CuO$_{4-y}$\cite{Hagen}; (k) the theoretical
result obtained from Eq.(3) with the exponent values of $m=-1$,
$\alpha=1.5$, $n=-0.3$, and $\beta=1.8$. }
\end{figure*}

{\it Universal Scaling}--- While the well-known puzzling scaling
relation $\rho_{xy}\propto\rho_{xx}^\beta$ remains controversial
\cite{Kang,Kopnin}, we find two universal scaling relations: \\
\begin{eqnarray}
\frac{\rho_{xy}}{\rho_m}\approx f(\frac{B-B_0}{B_m-B_0})\approx
f'\left(\frac{T-T_0}{T_m-T_0}\right) ,\\
\frac{\tan\theta_H}{{\tan\theta_H}^m}\approx
\Phi\left(\frac{B-B_0}{B_m-B_0}\right) \approx
\Phi'\left(\frac{T-T_0}{T_m-T_0}\right) .
\end{eqnarray}
this finding has been verified by experiments of variety of
different kinds of HTS's, from superclean YBCO \cite{Harris},
electron-doped superconductor
Nd$_{1.85}$Ce$_{0.15}$CuO$_{4-y}$\cite{Hagen} to HBCCO etc.
materials [6, 16-19, 22]. as shown in Fig1, where $\rho_m$ and
$\tan\theta_H^m$ are the first extreme values of Hall resistivity
and Hall angle from low temperature or low field sides. $B_m$ and
$T_m$ are the field and temperature corresponding to the extreme
point. $B_0$ and $T_0$ correspond to the specific field and
temperature where the Hall resistivity or $\tan\theta_H$ becomes
first measurable at the low temperature or low field side. k
represents the theoretical result obtained from Eq.(3) with the
exponent values of $m=-1$, $\alpha=1.5$, $n=-0.3$, and $\beta=1.8$.
From Fig.1, we find that functions $f$, $f'$, $\Phi$ and $\Phi'$ are
almost of the same form. It can be shown that these forms are
equivalent to the corresponding ones shown by Xu et al. for YBCO in
Ref. \cite{Xu}, though now this formulas can describe also the case
of superconductors with double sign reversal \cite{22}.

\vspace{5mm}

{\it Unified $\rho_{xy}$ Equation}--- Starting from the the
$\rho_{xy}$ equation shown in Ref. [14] as\\
\begin{eqnarray}
\rho_{xy}(T)
&=\frac{\beta_0B^2\rho_n}{\phi_0B_{c2}}\{(1+\overline{\gamma})\frac{\rho_{xx}^2}{\rho_f^2}\
-2\overline{\gamma}\frac{\rho_{xx}}{\rho_f}\}\\\nonumber
&=\frac{\beta_{0}\rho_{n}B^2}{\phi_0B_{c2}}
\{(1+\overline{\gamma})exp[-2\alpha\frac{B^m(1-\frac{T}{T_{c}})^\alpha}{\kappa
T}\\\nonumber
&*(1+\frac{J\rho_{xx}/\rho_{f}}{B^n(1-\frac{T}{T_{c}})^\beta}-\frac{J}{B^n(1-\frac{T}{T_{c}})^\beta})]
\\\nonumber &-2\overline{\gamma}exp[-\alpha\frac{B^m(1-\frac{T}{T_{c}})^\alpha}{\kappa
T}\\\nonumber
&*(1+\frac{J\rho_{xx}/\rho_{f}}{B^n(1-\frac{T}{T_{c}})^\beta}-\frac{J}{B^n(1-\frac{T}{T_{c}})^\beta})]\},
\end{eqnarray}
and considering the general relationship between the two dimensional
(2D) conductivities and resistivities ,we obtain a unified equation
of $\rho_{xy}$ which can give coherent understanding of the HTS's
behavior with and without double sign reversal. The microscopic
analysis of the flux flow Hall conductivity $\sigma_{xy}$ based on
TDGL\cite{KopninLopatin} show that
\begin{equation}
\sigma_{xy}=\sigma_{xy}^Q+\sigma_{xy}^V
\end{equation}
the first term describes the contribution arising from the
quasiparticles inside and around the vortex core. It has the same
sign as normal state and is proportional to $H$. The second term
represents the hydrodynamic contribution of vortex Hall
conductivity. Taking both these aspects into account, we find {\it
two unified equations} as the {\it natural extension} of Eq.(3): one
for fixed temperature (Eq.(5)) and the other for fixed magnetic
field (Eq.(6))
\begin{widetext}
\begin{equation}
\rho_{xy}=\frac{\beta_0B^2\rho_n}{\phi_0B_{c2}}\{(1+\overline{\gamma})\frac{\rho_{xx}^2}{\rho_f^2}\
-2\overline{\gamma}\frac{\rho_{xx}}{\rho_f}\}\{\frac{\sigma_{xy}^Q(T,B_m)+[\sigma_{xy}(T,B)-\sigma_{xy}^Q(T,B_m)]
\theta(B-B_m)}{\sigma_{xy}^Q(T,B_m)}\}
\end{equation}
\begin{equation}
\rho_{xy}=\frac{\beta_0B^2\rho_n}{\phi_0B_{c2}}\{(1+\overline{\gamma})\frac{\rho_{xx}^2}{\rho_f^2}\
-2\overline{\gamma}\frac{\rho_{xx}}{\rho_f}\}\{\frac{\sigma_{xy}^Q(T_m,B)+[\sigma_{xy}(T,B)-\sigma_{xy}^Q(T_m,B)]
\theta(T-T_m)}{\sigma_{xy}^Q(T_m,B)}\}
\end{equation}
\end{widetext}
where $\rho_f$ and $\rho_{xx}$ are the pinning free B--S flux flow
resistivity and the longitudinal nonlinear resistivity respectively.
Ref.[14] shows that the longitudinal $\rho_{xx}/\rho_f$ has the
extended power law form
\begin{equation}
\frac{\rho_{xx}}{\rho_f}\equiv\Omega=exp[-a\frac{U_c(B,T)}{kT}(1+(\Omega-1)\frac{J}{J_{c0}(B,T)})^p]
\end{equation}
This form agrees well with the well-known experimental data of
$\rho_{xx}$ by Koch et al.\cite{Koch} as well as the letter of
Strachan et al.\cite{Strachan} near the vortex glass transition (see
Ref.\cite{Ning}). The parameter $\overline{\gamma}$ is defined as
$\overline{\gamma}\equiv\gamma(1-\overline{H}/H_{c2})$ with
$\overline{H}$ the average magnetic field over the vortex core and
$\gamma$ is the contact force on the surface of vortex core
\cite{WDT,Xu}. $\beta_0\equiv\mu_mH_{c2}$ with $\mu_m=\tau e/m$ the
mobility of the charge carrier. The upper critical magnetic field is
given by $H_{c2}=\Phi_0/2\pi\xi^2$. $\theta(x)$ is the Heaviside
function. Eqs.(5) and (6) are not {\it confined only to the specific
case of YBCO }but able also to give a {\it full description of
superconductors with double sign reversal of $\rho_{xy}$}.
\\

{\it Doping Dependence problem}---Equation(5) provides a reasonable
explanation of the striking dopping dependence of Hall effect in the
superconducting state from the underdoped to overdoped regime of HTS
observed by Nagaoka et al. \cite{Nagaoka}. These authors found that
$\sigma_{xy}$ increases linearly with $H$ at high field in all
crystals. With decreasing $H$, $\sigma_{xy}$ diverges to $-\infty$
for underdoped crystals but diverges to $+\infty$ for overdoped
crystals. This tendency is consistent with our Eq.(5) though it
seems unlike that predicted by some theoretical models
\cite{Nagaoka,KopninLopatin}.

\vspace{3mm}

The comparison of our Fig.1 and Eqs.(1), (2) with Ref.[14] clearly
shows that the nonlinear $\rho_{xx}$ equation used by Xu.et al. for
YBCO is also effective for other superconductors. Moreover, since
$0<\gamma<1$ according to its definition \cite{WDT,Xu} and
$\rho_{xx}\ll\rho_f$ according to Eq.(7) at low field, one should
expect the negative sign of the left large bracket in Eq.(5) . This
remarkable feature naturally explains the results of Ref.[7], i.e.,
{\it at low field,the Hall sign is electron like in the underdoped
regime and hole like in the overdoped regime}, as shown also in our
Fig.1.
\\

{\it Discussion}---Besides the underdoped superclean YBCO and the
electron-doped Nd$_{1.85}$Ce$_{0.15}$CuO$_{4-y}$, some otherwise
HTS's like HBCCO etc. have double or triple sign reversal of Hall
effect. Kang et al. observed an scaling power law exponent
$\beta\approx1$ in HBCCO thin film with double Hall reversal and
columnar defects which consistent with a recent theory based on
d-wave superconductivity. Considering the gap models of d-wave
superconductors, Kopnin and Volovik (KV) show that the gapless
quasiparticles exist also outside far from the vortex cores and
under the influence of the vortex superflow velocity. For superclean
systems with $\triangle^2_\infty\tau/E_F\gg1$, this new aspect
results in the unusual magnetic field dependence of
conductivities\cite{28} where $\Delta_\infty/E_F$ is not very small
in HTS's, the superclean regime is accessible there. In such case,
at high field ($B>B_m$), the second large bracket at the right-hand
side of Eq.(5) provides a continuous extension of the $\rho_{xy}$
equation in Ref.[14] with consideration of the microscopic analysis
of $\sigma_{xy}$ by different models. Thus, on the basis of Eqs.(5)
and (6), a full comparison of Hall anomaly experiment with
prediction from different models can be made, which may open
promising perspectives for understanding the mechanisms of high
temperature superconductivity. Such kind full comparison of
Longitudinal resistivity transition experiment with theories does
show some interesting features. While the experimental data of
MgB$_2$ with parameters in agreement with the prediction of BCS
superconductivity, the parameters of experimental data of
high-quality untwinned optimally doped YBCO single crystal show
remarkable difference\cite{29}. The unusual properties of cuprates
may appear even more strinking in the underdoped region£¬such as the
observed large Nernst effect in an extended region above Tc in
hole-doped cuprates[30-34] and very recently, Docron-Leyraud et al.
unambiguously observed quantum oscillations and the Feimi surface in
an underdoped YBCO\cite{30}.
\\

{\it Conclusion}---We show universal scaling behavior of pinning
effect on the Hall resistivity $\rho_{xy}$ and Hall angle
$\theta_H$, in accord with this, we find a unified equation of
$\rho_{xy}$ for different superconductors which may explain the
striking doping dependence of Hall anomaly in superconducting state
found by Nakaoka et al.\cite{Nagaoka} and propose {\it a full
comparison of Hall anomaly experiment} with {\it predictions of
theoretical models}.
\\

This work is supported by the Ministry of Science \& Technology of
China (NKBRSG-G 1999064602) and the National Natural Science
Foundation of China under Grant No.10674005, No.10174003,
No.50377040 and No.90303008.


\begin{thebibliography}{99}
\bibitem{22} Denoting $B_{b}$ in Ref. [14] as our $B_m$ and the
single sign reversal $B_r$ as $B_1$ one finds that the scaling relation in Ref.[14] can be rewritten as\\
\begin{eqnarray}\nonumber
f(\frac{B-B_b}{B_r-B_b})\equiv f(\frac{B-B_m}{B_1-B_m})=f(
C1\frac{B-B_0}{B_m-B_0}-C2)\\
\nonumber with \hspace{4mm} C1=C2\equiv (B_m-B_0)/(B_1-B_m).
\end{eqnarray}
\end{thebibliography}
\end{document}